# RADIATION STABILITY OF NANOCOMPOSITE SCINTILLATORS

L. Prouzová Procházková[1,2], F. Hájek[2], M. Buryi[2,3], Z. Remeš[2], V. Čuba[1]

[1] Department of Nuclear Chemistry, Faculty of Nuclear Sciences and Physical Engineering, Czech Technical University in Prague, Břehová 7, 115 19 Prague 1, Czech Republic

[2] Institute of Physics of the Czech Academy of Sciences, Na Slovance 1999/2, 182 00, Prague 8, Czech Republic

[3] Institute of Plasma Physics of the Czech Academy of Sciences, U Slovanky 2525/1a, 182 00, Prague 8, Czech Republic

**Abstract**

Radiation hardness of scintillating nanocomposites consisting of inorganic scintillating nanocrystalline powders dispersed in organic matrices was studied under electron, X-ray and γ-ray irradiation. Samples including pure press-compacted pellets of powder ZnO:Ga and YSO:Ce, and the nanocomposites of powder ZnO:Ga and YSO:Ce embedded in polystyrene matrix with different fillings were investigated. Effects of radiation on radioluminescence and other optical properties of studied materials were evaluated. Bright burn effect related to nanocrystalline powder scintillators was observed at lower doses. Radiation damage in nanocomposite materials is related to the formation of radicals in polystyrene matrix. Extent of radiation damage decreases with ZnO:Ga filling. Presented results show the importance of systematic and complex study of the radiation stability of composite scintillators.

## 1. Introduction

Physical, chemical and mechanical stability in the field of ionizing radiation, also known as radiation hardness, are key parameters of all materials that are exposed to irradiation during their practical use. Moreover, radiation hardness is an essential parameter to assess the economy of scintillation detectors operation if high radiation stress is expected, and to evaluate the lifetime of such components in complex and expensive devices. The investigation of different types of radiation is also important from the perspective of high-energy physics (HEP), as materials used in calorimetry will be exposed to a combination of different types of ionizing radiation.

Currently, plastic and/or nanocomposite scintillating materials (composed of inorganic nanoparticles embedded in polymer matrices) are being considered for new generation of detectors in high energy physics and time-of-flight positron emission tomography (TOF PET) applications. While inorganic, usually bulk, scintillators tend to be considered advantageous for their radiation hardness, the main advantages of their plastic counterparts are low price, simplicity of processing, machining and shaping, as well as ease of replacement. These advantages are complemented with good luminescent properties and prospects for fast timing. However, organic scintillators feature relatively low radiation stability in the field of high energy radiation. Due to the advent of new nanocomposite scintillators developed for advanced applications, the radiation stability of these new inorganic/organic materials became also a serious concern.

Such nanocomposites consist of scintillating nanoparticles (i.e. particles whose size does not exceed 100 nm in at least one direction) embedded into the optical matrix. Nanoparticles of ZnO:Ga or $CsPbBr_3$ have a tunable luminescence and sub-nanosecond photoluminescence and scintillation decay [1,2]. Especially because of their superior timing properties, they were identified as promising materials for time-of-flight applications in medical imaging and high energy physics fields [3]. Regarding matrices, polystyrene (PS) was found to be an appropriate host material with many advantages. PS itself is a good scintillator and the use of appropriate incorporation method(s) allows for homogenous distribution of a scintillating nanopowder, which is a crucial requirement for effective nanocomposite. Moreover, the efficient non-radiative energy transfer from the host matrix towards ZnO:Ga was evidenced [2]. On the other hand, the low radiation hardness of PS can significantly limit their use in detection systems, where radiation damage of the matrix could affect the detection properties of such systems.

Radiation stability of both polymers and inorganic bulk scintillators has been systematically studied for several decades [4,5,6]. The predominant effects of radiation damage in bulk crystal scintillators are the radiation-induced absorption, or color center formation, rather than the loss of scintillation light yield [5]. Recently, a very detailed review on the radiation stability of polymer scintillators was published [7]. Nanocomposite scintillators combine the properties of polymer matrix and inorganic crystals and their behavior and change in parameters under irradiation has not yet been sufficiently mapped. We must consider several simultaneous events, such as radiation damage at the structural level, for example, breaking the polymer chain of matrix, or the creation of the optically active centers connected with induced absorption and creation or disappearance of emission bands and formation of free radicals. Also, reactions on the phase interface can occur in the case of nanocomposite materials. Moreover, the distribution of radiant energy in a nanocomposite will be very different from homogeneous inorganic or organic materials.

The mostly used polymers for scintillators are polystyrene (PS), polyvinyltoluene (PVT) and polymethylmethacrylate (PMMA). All three show different occurrences of individual types of damage and different production of gaseous products under γ-rays and accelerated electrons, as shown in [7,8]. PS seems to be one of the most resistant polymers from studied plastic scintillators [7] with a very low amount of gaseous products, whose production during irradiation can be very problematic for vacuum applications. Dominant radiation damage in PS should be formation of radicals and cross-linking [9]. These processes will depend on dose rate, total dose, the presence of oxygen and oxygen diffusion rate [6], recovery due to the environment and many other factors. In addition to the typical damage (macroradical formation, chain scission, cross-linking etc.), accelerated electrons can also cause discharge-related damage to the polymer component. For soft X-ray or electron radiation in electron microscopy, quantification of radiation damage to various types of polymers was performed, e.g. in [10].

This study is focused on the investigation of radiation stability of selected nanocomposite scintillators under X-rays and electron beam irradiation. We focused our research on the evaluation of radiation damage of various PS nanocomposites suitable as detectors for fast timing applications, namely ZnO:Ga-PS. The effect of X-rays and electron irradiation on the absorption and luminescence properties was investigated in detail.

## 2. Materials and Methods

### *2.1. Preparation of ZnO:Ga-PS samples*

Luminescent ZnO:$Ga^{3+}$ nanocrystalline powder was prepared via photo-induced precipitation, as described in [11]. Aqueous solution of $Zn^{2+}$ and $Ga^{3+}$ salts and hydrogen peroxide was irradiated by low-pressure mercury lamps (UV Technik Meyer GmbH; power input 25 W; emitting photons at 254 nm). Solid phase formed after UV irradiation was dried, annealed at 200 °C and further annealed at 1000°C in air in the Clasic 0415 VAC vacuum furnace and at 800 °C in the mixture of $Ar/H_2$ using the Setaram

LabSys Evo thermoanalyzer. This three-step annealing is necessary to reach good luminescence properties.

ZnO:Ga-polystyrene composites (ZnO:Ga-PS, Fig.1) were prepared in cooperation with NUVIA a.s. following the procedure previously reported in [12]. The amount of 0.5, 5 and 10 g of ZnO:Ga powder was mixed with 50 g of granulate of polystyrene (Synthos PS GP 171) in Brabender lab mixer to prepare the composite with 1wt%, 10wt% and 20wt% filling. The mixture was then press-compacted in stainless steel frame, which was placed between the pairs of glass and stainless-steel slabs, into the 1mm thick plate. Subsequently, a pellet of 25 mm in diameter was cut from the plate.

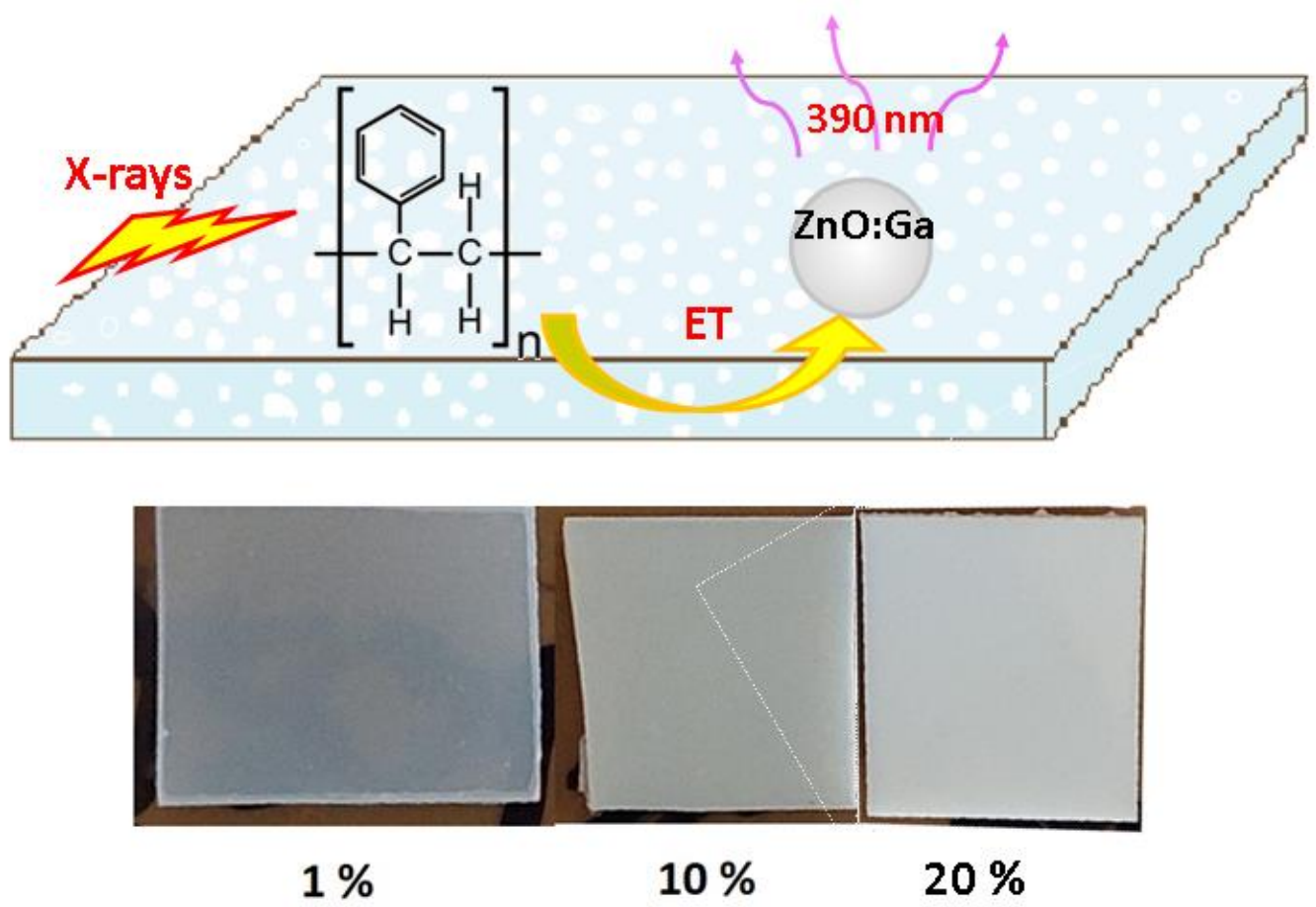

Fig.1: Scheme of ZnO:Ga-PS nanocomposite: non-radiative energy transfer between polystyrene matrix and ZnO:Ga; bottom – photographs of the nanocomposites investigated

*2.2. Irradiation methods*

Accelerated electrons irradiation was performed using high-frequency linear accelerator LINAC 4-1200 (Tesla V. T. Mikroel) with electron energy 4.5 MeV, pulse width 3 μs and repeating frequency 500 Hz. Samples were irradiated on Petri dishes placed on a mobile transporter. Dosimetry was established by alanine dosimeters. The deposition of energy in this case is very fast and highly non-uniform in short pulses (~3 μs), followed by long relaxation time (2 ms); calculated dose rate must be understood as an integral dose related to the total irradiation time.

A custom-made research X-ray cabinet SCIOX Beam was used for X-ray irradiation. The cabinet is equipped with two wide-angle tungsten low and high-voltage X-ray tubes, an oil-based cooling system, and a lead-shielded irradiation chamber 750 × 750 × 720 mm. The parameters of the irradiation were 195 kV, 20 mA. The distance between X-ray tube and irradiated samples was fixed at 400 mm. Dose rate was estimated for this configuration using the Fricke dosimeter.

Gammacell 220 (AECL Canada Ltd., dose rate evaluated at the time of irradiation ~14 $Gy \cdot h^{-1}$) was used as a radionuclide source of γ-rays. Radionuclide $^{60}Co$ is placed around the irradiated cell in the form of thin rods. Cell is manipulated by the control desk on the apparatus and shielded by the massive lead container. The dose rate was evaluated using Fricke dosimeter.

*2.3. Characterization*

Luminescent properties were evaluated by measuring radioluminescence emission spectra (RL) using the custom-made spectrofluorometer 5000 M, Horiba Jobin Yvon, equipped with the X-ray tube (RL spectra). The detection part of the set-up consists of single grating monochromator and photon counting detector TBX-04. Measured spectra were corrected for the spectral dependence of detection sensitivity. We compare RL spectra of our samples with radioluminescence BGO standard ($Bi_4Ge_3O_{12}$) in powder form. For further details see [12].

Cathodoluminescence (CL) measurements were performed in Scanning Electron Microscope Philips XL30ESEM. The acceleration voltage 30 kV and current 30 nA was used. CL spectra were collected using home-built optical setup consisting of parabolic mirror, optical fiber and spectrometer AvaSpec ULS2048LTEC. The spectra were corrected for instrumental sensitivity. Prior to the measurements, the sample surface was metalized (10 nm of Ti) to prevent surface charging. The measurement was done with the defocused spot (diameter 140 μm) fixed at one position. CL spectra were recorded every 10 s to observe the degradation process. Following the calculation in [13], the dose rate is 19.5 $MGy \cdot s^{-1}$ (assuming that the composite scintillator can be approximated by its matrix in the calculations).

Electron paramagnetic resonance (EPR) spectra of X-band (9.4 GHz) were obtained on a Bruker EMXplus spectrometer at the temperatures below liquid nitrogen temperature using Oxford Instruments cryostat ESR 900 and liquid helium cooling. The spectrometer sensitivity is about $10^{12}$ spins/mT, while microwave power of 2 mW, modulation amplitude of 0.1–0.5 mT and time constant of 160 μs were used.

The dry-air ventilated Nicolet IS50 spectrometer equipped with transmission (T) and Raman accessory was used for Fourier Transform Infrared spectroscopy (FTIR) measurements. T spectra were collected in 400-4000 $cm^{-1}$ spectral range with 4 $cm^{-1}$ spectral resolution using a heat source, KBr beamsplitter and DTGS detector. Raman measurements were collected in 70-3500 $cm^{-1}$ spectral range with 2 $cm^{-1}$ resolution using up to 500 mW YAG:Nd laser with excitation at 1064 nm and InGaAs photodiode detector.

The transmittance, reflectance and absorptance spectra were measured simultaneously in the 300–1400 nm spectral range by photothermal deflection spectroscopy (PDS) setup with 150 W Xe lamp, SpectraPro-150 monochromator (150-mm focal length, f/4-aperture, slits 1/1mm) equipped with two gratings: a UV holographic (1200 /mm) and a ruled (600 /mm) blazed at 500 nm. The spectral resolution was 5 nm with the UV holographic grating and 10 nm with the ruled grating. Samples were immersed into liquid (Florinert FC72) to measure the relative temperature of the illuminated sample independently for selected photon energies using deflection of probe laser beam. The spectra were spectrally calibrated by measuring PDS of a black carbon sample [14,15].

## 3. Results and discussion

Samples were irradiated by different sources of radiation and their optical parameters were studied – especially radioluminescence and absorption spectra. Considering that dose rate has a crucial effect on the radiation damage of polymers [4,7], we used four different sources – two accelerated electron sources with very high dose rates, X-ray source with a moderate dose rate, and γ source with a low dose rate; dose rates and the evaluated of total doses received are listed in Tab. 1.

Tab.1: Parameters of radiation used in this study

| Type of radiation | Source | Energy | Dose rate | Total dose |
|---|---|---|---|---|
| $e^-$ | Linear accelerator | 4 MeV | 200 $kGy \cdot h^{-1}$ | 600 kGy |
| | Electron microscopy | 30 keV | 19.5 $MGy \cdot s^{-1}$ | 19 500 MGy |
| γ-rays | Gammacell | 1.25 MeV | 14 $Gy \cdot h^{-1}$ | 15 kGy |
| X-rays | Laboratory X-ray irradiator | 72 keV | 300 Gy·h$^{-1}$ | 7.5 kGy |

### *3.1. Irradiation by accelerated electrons*

#### *3.1.1. Linear accelerator*

Samples in the form of 1×1cm plates (1mm thickness) were irradiated by accelerated electrons from linear accelerator to receive doses 200, 400 and 600 kGy. The heating of samples during irradiation was not significant and was not monitored in detail. A slight yellow-brown coloration was observed with increasing dose. Samples were characterized by RL, FTIR and EPR within 48 hours of irradiation to minimize the influence of healing on the measured properties.

In RL spectra (Fig. 2a,b), we observe the change in the intensity of characteristic exciton-related emission peak with maximum at 390 nm, which we attribute to the ZnO:Ga grains in the nanocomposite. The mechanism of energy transfer from polystyrene matrix and ZnO grains which we discussed already in [12] probably plays a significant role in degradation of luminescence properties. Wide emission band in Fig. 1b is characteristic for defect-related ZnO luminescence and is probably caused by surface defects on the grains formed during composite preparation (usually defect-related luminescence is completely suppressed in the nanopowder, as shown in our previous work in [11,12]). A more significant decrease in RL intensity was observed for samples with lower filling with ZnO:Ga particles. At lower doses (100 kGy), so-called bright burn [16,17,18], which means radiation-induced enhancement of luminescence, was observed for 1% filling (Fig. 2a). At higher doses, the deterioration of luminescence properties was observed, which was practically negligible in the sample with 10% filling (Fig. 2b). For further characterization, we used samples irradiated by the higher dose (400 kGy for EPR and 600 kGy for absorptance and FTIR).

The bright burn or sensitization effect is well described for some bulk inorganic scintillators, such as YAG:Ce or LSO:Ce [16,17] and has been ascribed to a progressive electron trap filling, increasing scintillation efficiency. We also studied the effect of radiation on the powdered scintillator YSO:Ce in pellet form (Fig. 3a) and in the form of a composite (YSO:Ce-PS 1%, Fig. 3b) to distinguish this phenomenon from structural changes at the level of chemical bonds in the matrix leading to the improvement of energy transfer from PS to the scintillating grains. We can clearly see that RL intensity is strongly enhanced after irradiation with a dose of 100 kGy in both cases. Further irradiation leads to the deterioration of luminescence properties. Next to the main emission band characteristic of the used inorganic scintillators, emission band around 320 nm is recognizable in RL spectra (Fig. 2a and Fig. 3b) for composites with low filling. This band belongs to the PS matrix and decreases gradually with increasing dose. It is also worth noting that the shape and position of emission band slightly changes after irradiation in the case of YSO:Ce-based samples, probably due to the filling of high-temperature stable charge carrier traps [16]. This effect has not yet been sufficiently explained in the literature. Detailed studies, e.g. using EPR, are rather rare. For majority of the applications, this enhanced emission is more of a complication than a benefit, as the detectors usually require the greatest possible response stability over a long-time horizon [17].

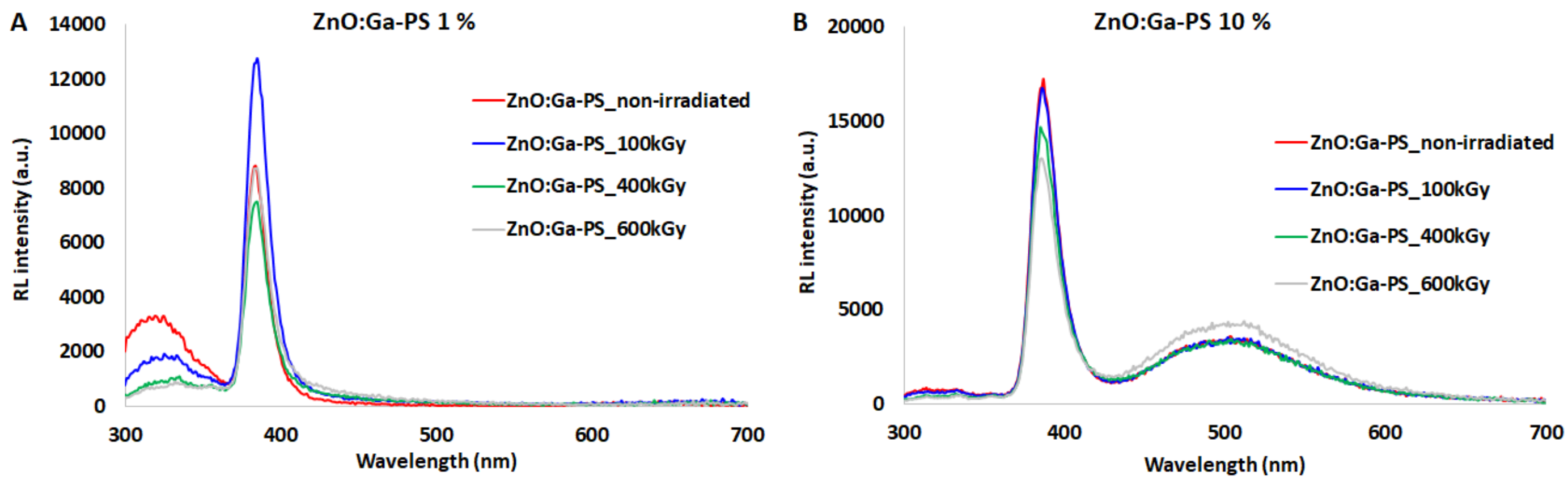

Fig. 2a,b: RL spectra of ZnO:Ga-PS 1 % (a) and 10 % (b) samples irradiated under $e^-$ beam

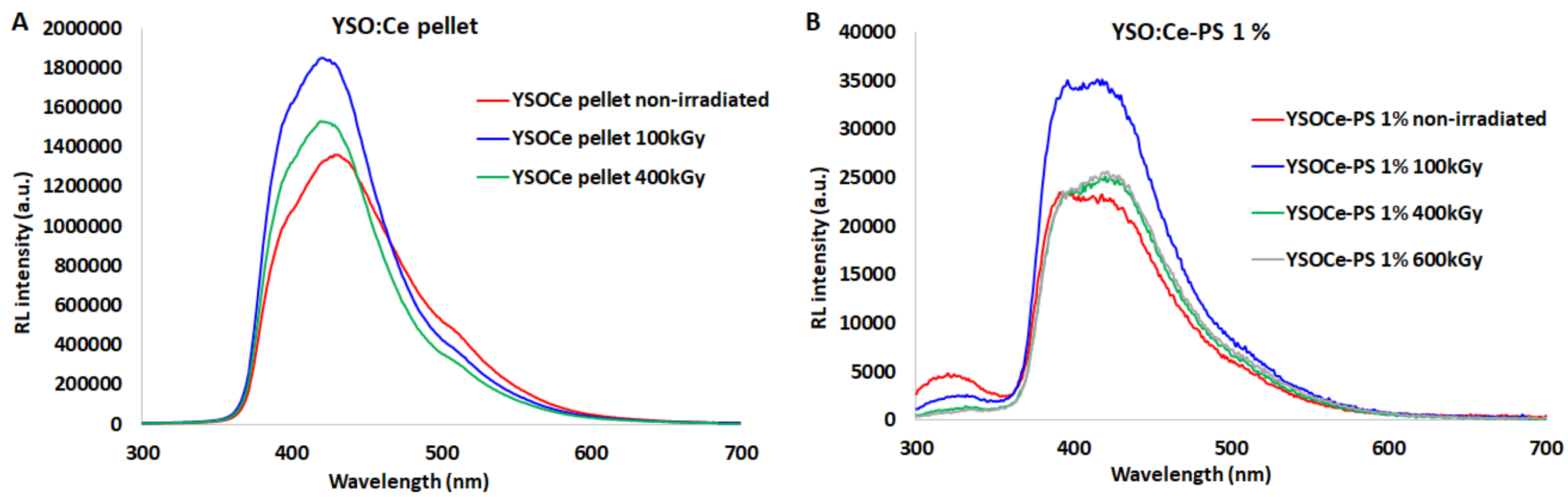

Fig. 3a,b: RL spectra of YSO:Ce pellet (a) and YSO:Ce-PS 1 % (b) irradiated under $e^-$ beam

EPR spectra measured in several types of samples are shown in Fig. 4. The pure polystyrene sample (PS, AG) demonstrated no significant signals. Electron beam irradiation (the dose was 400 kGy) causes the appearance of a relatively intense new signal (S1). It is characterized by three principal values of the *g* tensor: $g_1$ = 2.017 (3325 G), $g_2$ = 2.002 (3350 G), $g_3$ = 1.991 (3370 G). This signal is unique, as these values do not appear in the other papers dedicated to the high-dose irradiated polystyrene (see e.g., [19-21]). It should be noted that there were different *g*-factors reported in the above-mentioned papers. However, the origin of these signals was expected to be different oxygen-based radicals. The same conclusion can be made for the origin of S1 signal.

The as grown ZnO:Ga nanorods were reported to feature the shallow-donor signal (SD in Fig. 4) [22]. However, it is totally absent in ZnO:Ga presented in this study, irradiated with the electron beam with the dose of 400 kGy (Fig. 4). One may thus conclude that considering the origin of the SD signal ($Zn^+$ + D; D = Al, Ga or H) [23,24], the following hole trapping mechanism can be expected:

$Zn^{+} + D + h^{+} \xrightarrow{el.\ beam} Zn^{2+} + D$ [25,26]

$Zn^{2+}$ is not paramagnetic. This can be confirmed by measuring the ZnO:Ga embedded into the polystyrene as grown (ZnO:Ga-PS (20%), AG in Fig. 4) and irradiated with the electron beam imparting the dose of 400 kGy (ZnO:Ga-PS (20%), 400 kGy in Fig. 4) to the samples. The drop in the SD signal intensity (about 1.5 times, not 100%) has been observed after the irradiation. Interestingly, the weak drop of the S1 in the ZnO:Ga-PS (20%), 400 kGy sample has been observed as compared to the PS, AG one indicating that the charge trapping processes appear also in the polystyrene as well.

These results clearly show that the only radicals present in irradiated nanocomposite originate from PS matrix, rather than ZnO:Ga. The kinetics of radical product accumulation can affect the destruction of chain bonds (scission of the main chain) or the formation of new bonds (cross-linking), but gaseous products can also be formed or oxygen from the environment can be bound by the resulting radicals. All structural changes may deteriorate scintillating properties.

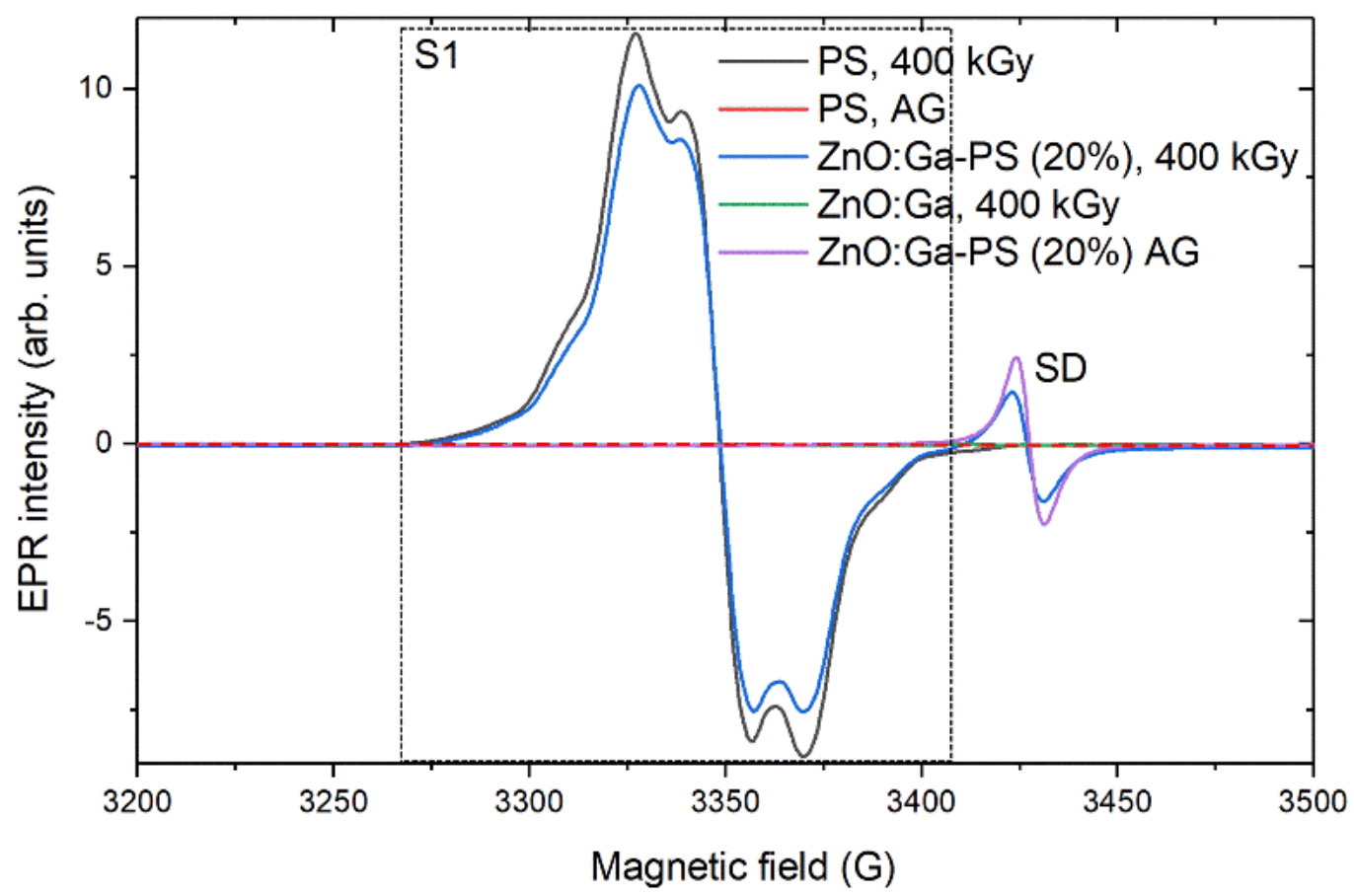

Fig. 4: EPR spectra of the PS, AG (after grow) and 400 kGy irradiated; ZnO:Ga, 400 KGy irradiated; ZnO:Ga-PS (20%), as grown and 400 kGy irradiated, measured at microwave frequency f = 9389 MHz and room temperature. The S1 signal (stressed by the dashed oblong) originating from PS and the shallow donor (SD) signal from the ZnO structure are indicated.

### *3.1.2. Cathodoluminescence*

CL spectra/time maps for samples ZnO: Ga PS 10% and ZnO:Ga PS 1% are shown in Fig 5a and 5b, respectively. The maxima at 3.2 eV (387 nm) – band 1, 3.7 eV (335 nm) – band 2 and 2.35 eV (527 nm) – band 3, are assigned to ZnO excitonic emission, PS matrix emission, and an emission from defects in ZnO, respectively. The intensity of all luminescence bands decreased with increasing the applied dose, however, the characteristic dose, at which the luminescence decrease to 1/e is significantly lower for PS matrix (22 MGy) than for emission bands associated with ZnO nanoparticles for both samples. These characteristic doses are 3240 and 1800 MGy for excitonic emission, and 3240 and 2460 MGy for defect-related luminescence for samples with 10 and 1 wt. % loading, respectively. Note, however, that in the initial stage of the irradiation (till dose 58.5 and 39 MGy for the respected samples), the luminescence of the defect emission is enhanced. This could be done i.e. by defect activation by a hydrogen dissociation from the defect or filling the minority carrier traps in the material. The different stability of ZnO:Ga and PS matrix is given by the nature of the components – ZnO:Ga inorganic crystals are less susceptible to radiation induced degradation in comparison with polymers, also due to the number of chemical bonds and radicals formed under irradiation.

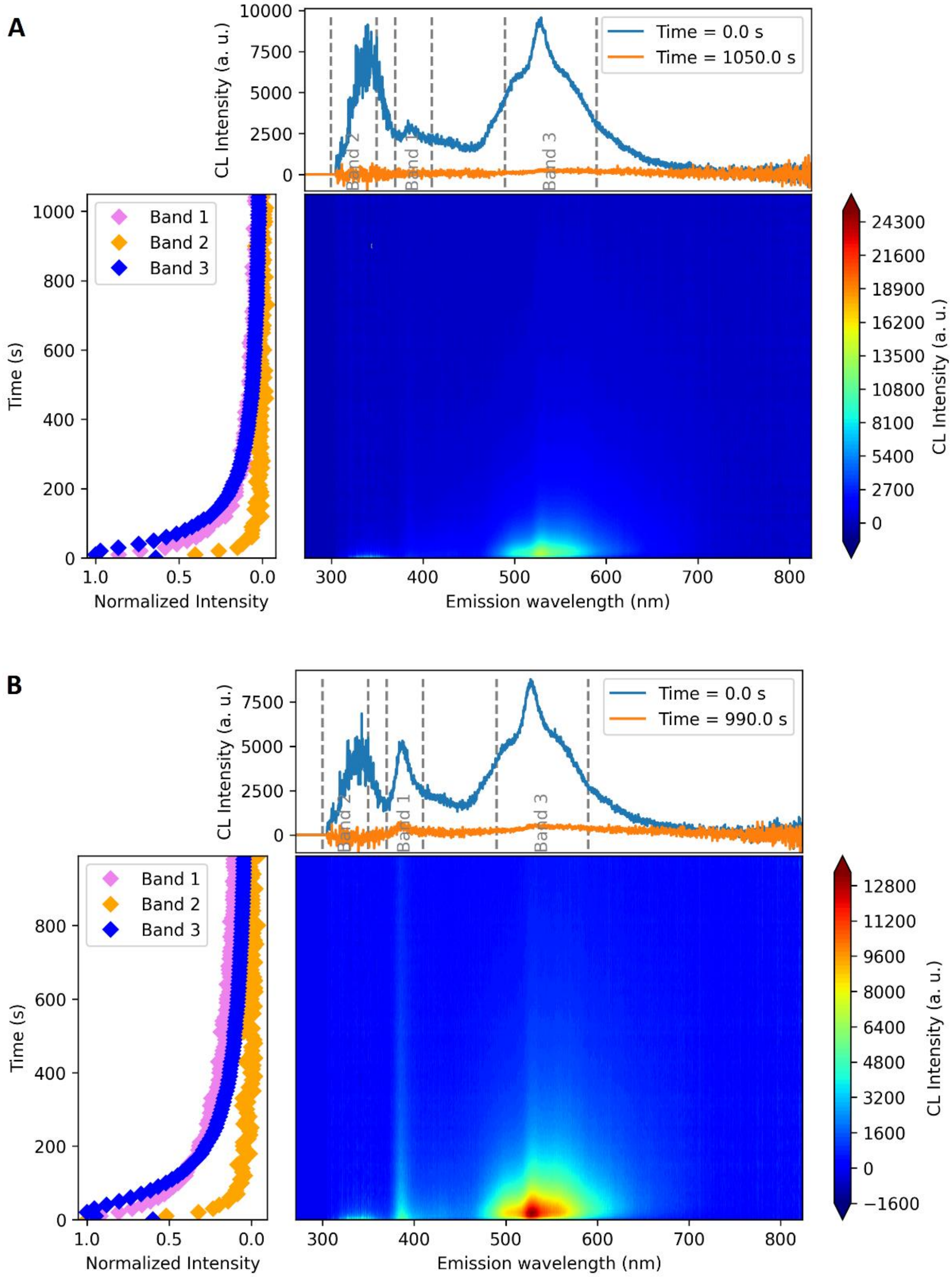

Fig. 5a,b: CL time-resolved spectra for ZnO:Ga-PS 10 % (a) and ZnO:Ga-PS 1 % (b)

### *3.2. Irradiation by X-rays, γ rays*

Irradiation by indirectly ionizing radiation was performed using X-rays with mean energy of about 72 keV and γ-rays with mean energy 1.25 MeV. The aim was to compare effect of relatively low-energy photons and high-energy photons with different dose rates. We suppose that X-rays and γ-rays should not have as high degradation effect as accelerated electrons [7]. Samples irradiated by γ-rays were measured when they reached the dose 6 kGy, 10 kGy and 15 kGy. Pure ZnO:Ga pellet pressed from ZnO:Ga nanocrystalline powder was also irradiated to compare the effect on the scintillating powder and polystyrene nanocomposite, given the results obtained for electron irradiation, which show the presence of the bright burn phenomenon. This phenomenon was confirmed for ZnO:Ga pellet (Fig. 6a) and it is clearly evident also for composite samples with low filling (ZnO:Ga-PS 1 %, see Fig. 6b). This enhancement of luminescence intensity is not so strong in the case of samples with higher filling (ZnO:Ga 10 %, see Fig 6c) and it is suppressed with higher doses, probably due to the degradation of polystyrene matrix. To better explain the observed trend, it would be appropriate to measure the dependence of luminescence on a larger range of doses.

X-rays irradiation was carried out as intermittent irradiation in the mode of 2 hours of irradiation - 1 hour of cooling of the irradiation box to prevent overheating of the X-ray lamp. Samples were characterized after reaching total dose ca. 7.5 kGy (Fig. 6b,c).

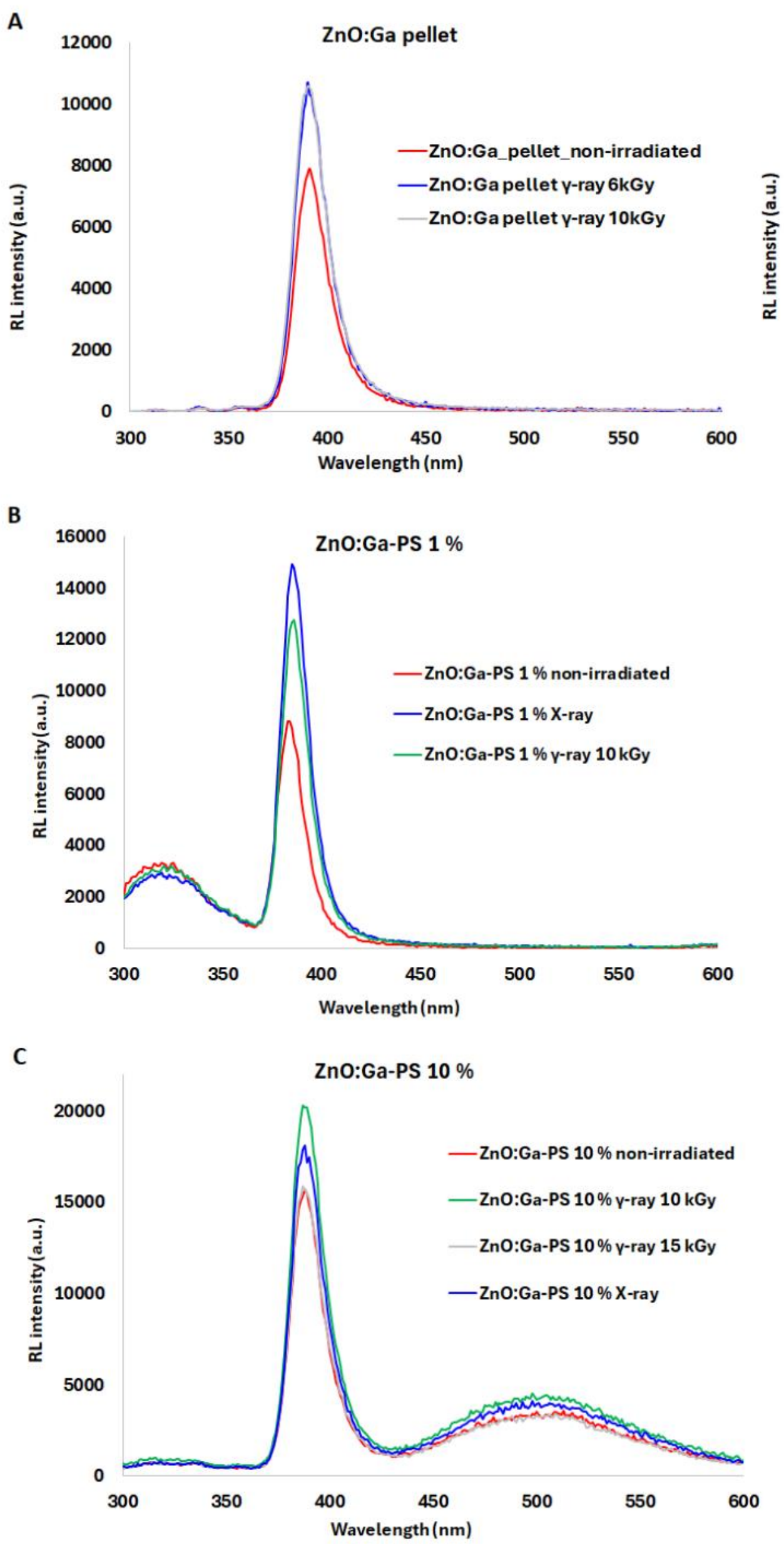

Fig. 6a,b,c: RL spectra of ZnO:Ga powder-pressed pellets irradiated under γ radiation (a), RL spectra of ZnO:Ga-PS 1 % (b) and 10 % (c) samples irradiated under X-rays and γ radiation

The optical properties of the nanocomposite samples irradiated under accelerated electrons and X-rays were compared using absorptance (Fig. 7), FTIR transmission (Fig. 8) and Raman spectroscopy (Fig. 9) measurements, which were only measured for the 1% filling composite due to the transparency of the sample. Significant changes near the optical absorption edge were observed in violet and near UV optical absorptance after electron beam irradiation (Fig. 7), which was also associated with the change in the sample color to yellow, while no measurable changes were recorded by FTIR and Raman measurement.

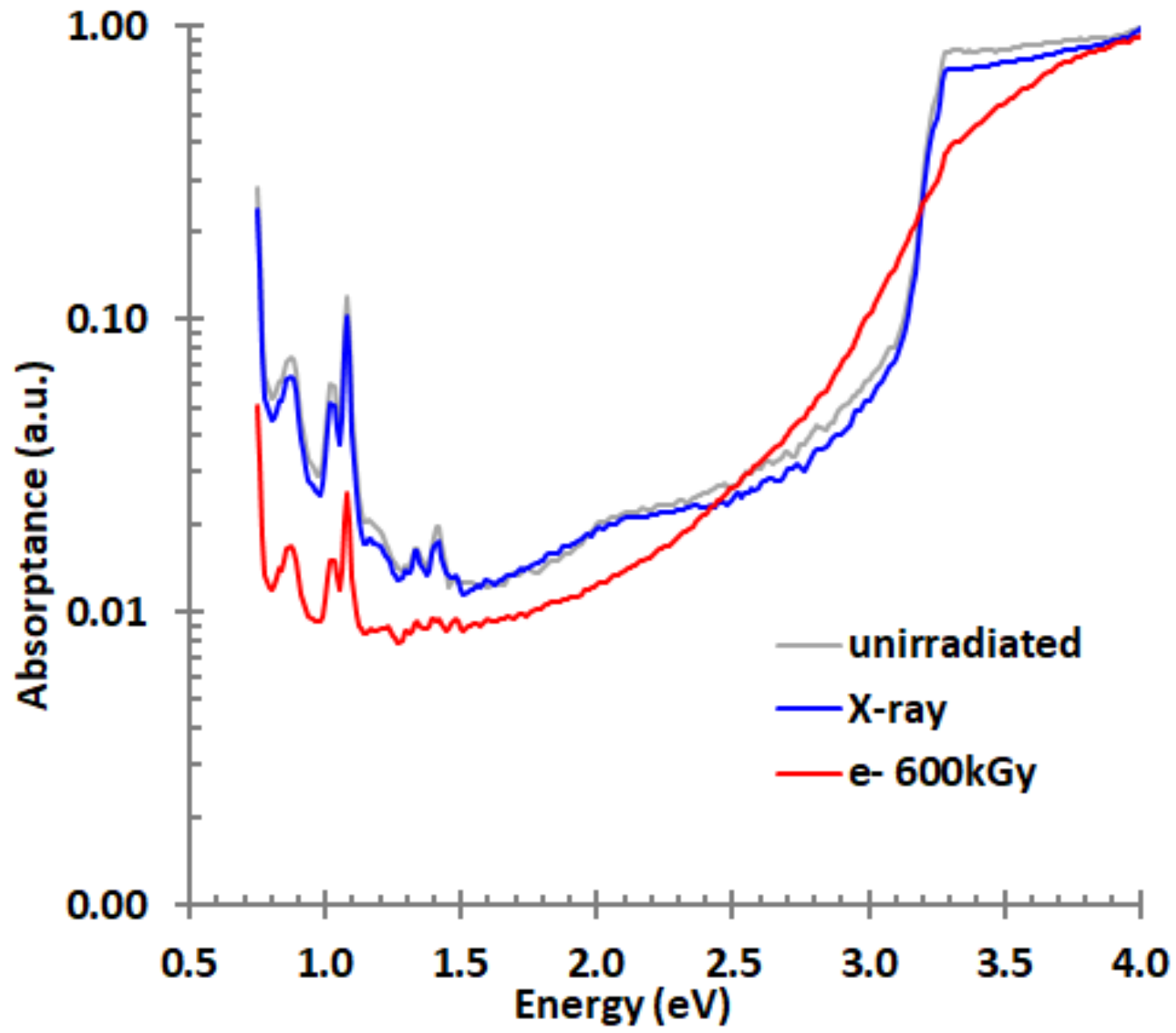

Fig. 7: Absorptance spectra measured for irradiated samples of ZnO:Ga-PS 1 %

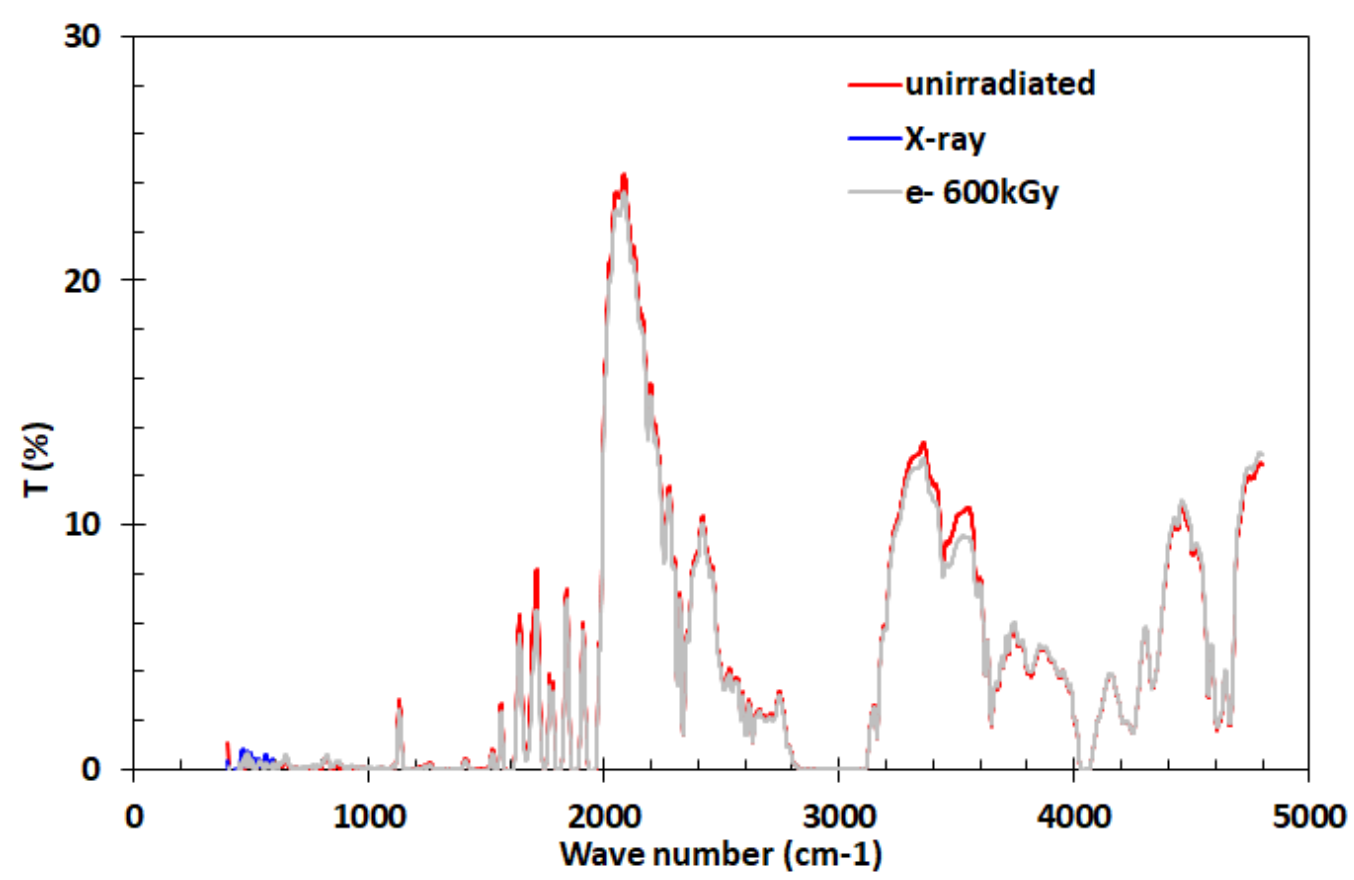

Fig 8: FTIR transmission spectra measured for irradiated samples of ZnO:Ga-PS 1 %

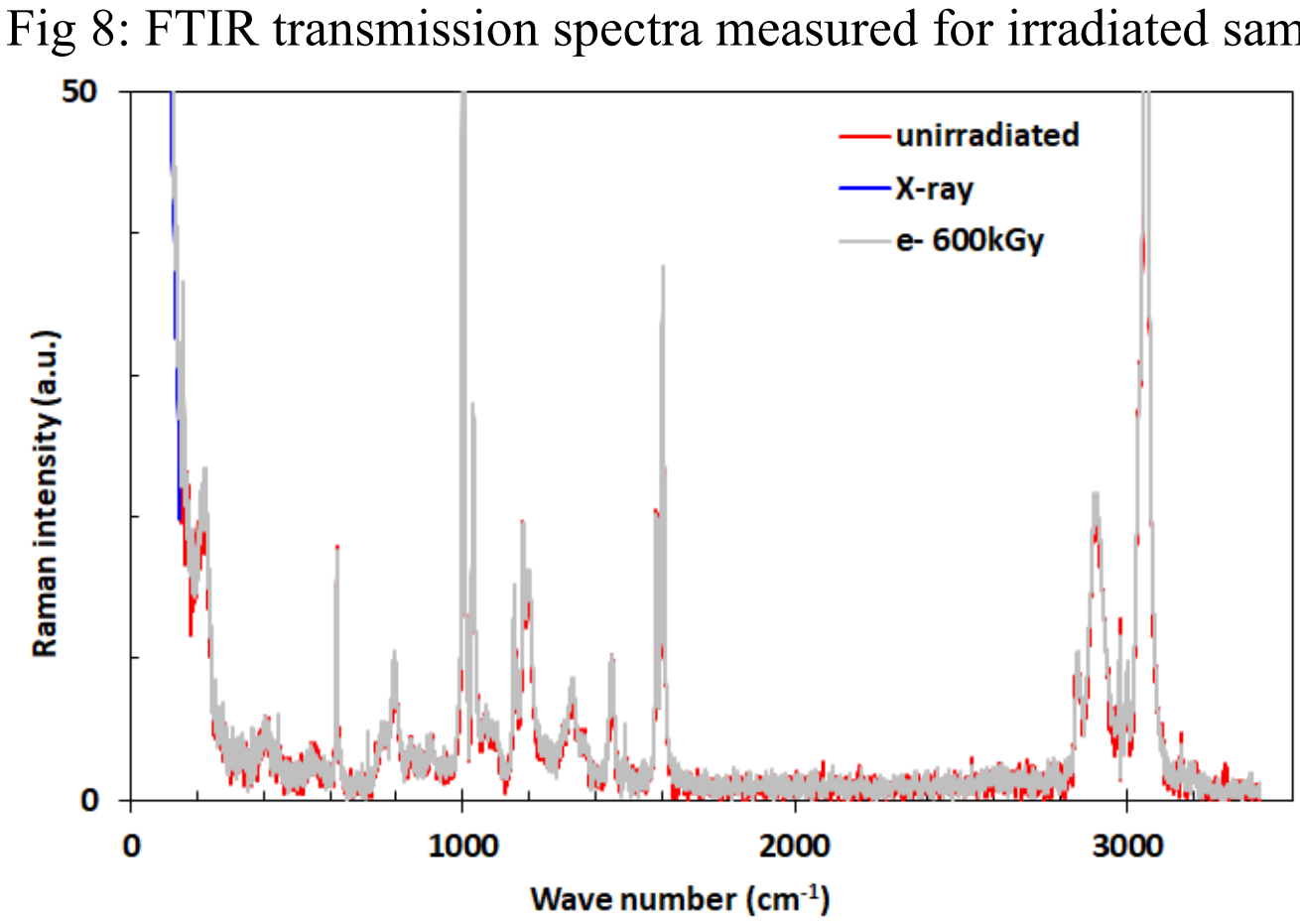

Fig. 9: FTIR Raman spectra measured for irradiated samples of ZnO:Ga-PS 1 %

## 4. Conclusions

Presented experiments are based on the electron, X-rays and γ-rays irradiation followed by characterization of radioluminescence and optical properties. Radioluminescence and absorption spectra can show even very small changes caused either at the structural level in the matrix or by changes in the electronic structure of scintillation crystals – free radicals formation or defects manifesting as electron traps or optically active centers, etc. Furthermore, stability of scintillation characteristics is one of the basic requirements for modern scintillation materials.

The presented results lead to the following conclusions: electron irradiation causes the deterioration of the optical parameters, such as absorbance and radioluminescence intensity, while X-rays and γ-rays do not negatively affect ZnO:Ga-PS samples in studied dose ranges. Bright burn effect, so far typically observed in bulk inorganic crystals, was observed for press-compacted pellets of scintillating powders as well as for nanocomposite materials; this effect was stronger in the case of samples with lower powder filling. ZnO:Ga-based samples retained its luminescence parameters, especially the strong exciton-related emission enabled by its nanocrystalline nature.

EPR measurements indicated the creation of oxygen-based radicals in polystyrene matrix. Changes in charge delocalization or local radical disorders may also occur in the matrix. Nanocomposites with higher powder filling show higher stability of studied properties.

Expected doses received during the LHC operation period are supposed in the range 10-100 kGy [7], but the dose rate ranges will be crucial, as they are essential for assessing the extent of radiation damage. In medical imaging, dose rates and total doses are usually very low. It can be assumed that the studied materials for TOF PET would be sufficiently stable throughout the entire operation period, as the dose range studied was several orders of magnitude higher.

In a follow-up study we will quantitatively determine the degradation of scintillators by light yields and afterglows measurements. Note that in the presented study, relatively low doses at various rates were received by the samples irradiated with a particular type of radiation. The future study will focus on the higher doses received at comparable dose rates for the respected types of radiation to complete the picture of degradation for nanocomposite scintillators in PS matrix.

Presented results show the importance of systematic and complex study of the radiation stability of composite scintillators. A deeper study of property changes and understanding of the radiation distribution mechanism in nanocomposites could enable the development of new materials with improved radiation resistance.

**Acknowledgement**

This research has been supported by OP JAC financed by ESIF and the MEYS (Project No. LASCIMAT - CZ.02.01.01/00/23_020/0008525) and CSF project GC24-10607J.

**CRediT authorship contribution statement**

**Lenka Prouzová Procházková**: Investigation, Methodology, Writing – original draft, Visualization; **František Hájek**: Investigation; **Maksym Buryi**: Investigation; **Zdeněk Remeš**: Investigation; **Václav Čuba**: Supervision, Writing – review & editing

**Declaration of Competing Interest**

The authors declare no conflict of interest. The funders had no role in the design of the study; in the collection, analyses, or interpretation of data; in the writing of the manuscript, or in the decision to publish the results.

**Data availability**

Data will be made available on request.